\documentclass[journal]{IEEEtran}

\usepackage{graphicx}
\usepackage{amssymb,amsmath}
\usepackage{multirow}
\usepackage{tabularx}
\usepackage{acronym}
\usepackage{slashbox}
\usepackage{color}
\usepackage{stfloats}
\usepackage{lettrine}
\usepackage{mdwlist}
\usepackage{cite}

\begin{document}

\title{Performance Analysis of Non-DC-Biased OFDM}
\author{
  Yichen Li\authorrefmark{1}, Dobroslav Tsonev\authorrefmark{1} and
Harald Haas\authorrefmark{1}\\
 \begin{minipage}{8cm}
   \begin{center}
     \vspace*{0.3cm}
  {\authorrefmark{1}\normalsize\sl Institute for Digital Communications}\\
  {\normalsize\sl  Joint Research Institute for Signal and Image
Processing}\\
  {\normalsize\sl  School of Engineering}\\
  {\normalsize\sl  The University of Edinburgh}\\
  {\normalsize\sl EH9 3JL, Edinburgh, UK}\\
  {\normalsize   \{yichen.li, d.tsonev, h.haas\}@ed.ac.uk
\vspace*{0mm}}\vspace*{0.0cm}
   \end{center}
   \end{minipage}
         \hspace*{0.0cm}
}

\maketitle

\begin{abstract}

The performance analysis of a novel optical modulation scheme is presented in this paper. The basic concept is to transmit signs of modulated optical orthogonal frequency division multiplexing (O-OFDM) symbols and absolute values of the symbols separately by two information carrying units: 1) indices of two light emitting diodes (LED) transmitters that represent positive and negative signs separately; and 2) optical intensity symbols that carry the absolute values of signals. The new approach, referred as to non-DC-biased OFDM (NDC-OFDM), uses the optical spatial modulation (OSM) technique to eliminate the effect of the clipping distortion in DC-biased optical OFDM (DCO-OFDM). In addition, it can achieve similar advantages as the conventional unipolar modulation scheme, asymmetrically clipped optical OFDM (ACO-OFDM), without using additional subcarriers. In this paper, the analytical BER performance is compared with the Monte Carlo result in order to prove the reliability of the new method. Moreover, the practical BER performance of NDC-OFDM with DCO-OFDM and ACO-OFDM is compared for different constellation sizes to verify the improvement of NDC-OFDM on the spectral and power efficiencies.

\end{abstract}


\section{Introduction}

\lettrine[lines=2]{\textbf{W}}{ITH} the rapid development of wireless services and applications, since 2000 wireless data rates have been growing exponentially. Some recent forecasts indicate that $5^{th}$ generation (5G) wireless systems will have speeds of 1~Gbps by 2020 \cite{5G2013}. Despite the fact that the hardware of the system can satisfy the requirement of the high speed transmission, the limited radio frequency (RF) spectrum may not be sufficient to cope with future data rate demands. As a viable complementary approach, optical wireless communication (OWC) has gained significant attention in part due to recent technological developments in solid state lighting technology \cite{emh1101}. The momentous advantage of OWC is that it offers almost infinite bandwidth ranging from infrared (IR) to ultraviolet (UV) including the visible light spectrum \cite{mmeh1001}. Other important benefits of OWC are: license-free operation; high communication security; low-cost-front-ends; and no interference to RF systems meaning that OWC and RF systems can be used simultaneously.

In current visible light communication (VLC) systems, high speed light emitting diodes (LEDs) are mainly used as transmitters. At the receiver, highly sensitive photodiodes (PDs), such as positive-intrinsic-negative (PIN) diodes, avalanche photo diodes (APDs) and single-photon avalanche diodes (SPADs) are used \cite{SPAD2013}. To date, the fastest wireless VLC system using a single LED can achieve speeds exceeding 3 Gb/s \cite{TD2013}. However, the incoherent light output of the LED means that information can only be encoded in the intensity level. As a consequence, only real-valued and positive signals can be used for data modulation. This is in contrast to RF systems which make use of complex valued and bi-polar signals. Thus, VLC systems are usually considered to be modulated as an intensity modulation (IM) and direct detection (DD) system \cite{kb9701}. On-off keying (OOK), pulse position modulation (PPM) and pulse amplitude modulation (PAM) are some of the common modulation schemes used in conjunction with IM/DD systems \cite{kb9701,mz0601,wbcb0501,lrbk0901}. Recently, the Optical Orthogonal Frequency Division Multiplexing (O-OFDM) modulation scheme, treated as the high-speed data transmission approach, can also applied in the context of IM/DD systems \cite{a0901,meh1001,emhp0701,mhh1201,lvja1201}.


\subsection{Optical OFDM}

For the high-speed OWC system, O-OFDM is used to handle severe ISI. The advantages of OFDM in OWC are same as in RF which are described in \cite{a0901,c0501}. However, as the O-OFDM is based on the IM/DD system which is limited to transmit real-valued signals, the set-up methods of each subcarrier are different between O-OFDM and the traditional OFDM system in RF. In O-OFDM, during the signal generation phase, real-valued symbols can be achieved by imposing Hermitian symmetry on the information frame before the inverse fast Fourier transform (IFFT) operation. This decreases the spectral efficiency by half. In general, standard techniques to ensure positive optical signals are DC-biased optical OFDM (DCO-OFDM), asymmetrically clipped optical OFDM (ACO-OFDM) and unipolar OFDM (U-OFDM) \cite{as0801,tsh1201}. In DCO-OFDM, a DC-bias is added to the original OFDM signal and the negative part is clipped. Clipping in DCO-OFDM may cause nonlinear distortion \cite{dsh1102}. If the DC-bias is increased to an optimal level, all of the symbols will be positive but the higher level requires more transmission power. In ACO-OFDM, the system inserts zeros on even subcarriers and modulates only odd subcarriers. As a result, a group of antisymmetric real-valued OFDM symbols are obtained, as shown in \cite{dsh1102}. This allows any negative samples to be clipped without distortion. Since only half of the subcarriers carry information bits, the spectral efficiency of ACO-OFDM is about half the spectral efficiency of DCO-OFDM. In U-OFDM, the positive part of OFDM symbols and the negative part of the symbols will be transmitted respectively \cite{tsh1201}. The positive block comes from the original OFDM signal with clipping the negative part and the negative block is generated in the same way. At the transmitter, the positive block is transmitted first and the absolute value of the negative block is then transmitted. Since the length of the OFDM frame is doubled, U-OFDM has the same spectral efficiency as ACO-OFDM.

\subsection{Optical Spatial Modulation}

In current 4G communication systems, OFDM multiple-input multiple-output (MIMO) is used as an efficient and effective method to satisfy the demand of high data rate transmission without ISI \cite{book:t0901,osjwfo0701,tech:m0703}. Examples of MIMO techniques are vertical Bell Labs layered space-time (V-BLAST), Alamouti and spatial modulation (SM) \cite{wfgv9801,mhsay0801,yxzl1101}. Compared to V-BLAST and Alamouti, SM has better BER performance while achieving the same spectral efficiency. In addition, about $90\%$ reduction in receiver complexity can be achieved \cite{mhsay0801}. For the VLC system, the optical SM (OSM) technique using IM/DD has been considered in \cite{meh1101}. In OSM, within a room, multiple transmitters are spatially separated. Only one LED is activated at any time instance and visible light is emitted with a fixed frequency and a certain optical power. Each transmitter index carries $\log_2(N_t)$ bits when the number of transmitters is $N_t$. In the IM/DD system, the value of modulated signals can be transmitted by the certain optical power. Thus, the conventional modulation schemes can be used in OSM, and even O-OFDM. In the conventional OSM-OFDM system, the indices of the transmitters carry a part of information bits, and modulated signals, which carry the other part of information bits, are transmitted by the active LED. The detailed model of the conventional OSM-OFDM system has been introduced in \cite{ydh1301}.

\subsection{NDC-OFDM}

A novel O-OFDM modulation scheme designed for OSM is presented in \cite{ydh1301}, which combines the basic OSM-OFDM and the original O-OFDM modulator, referred as to Non-DC-biased OFDM (NDC-OFDM). The new method aims to eliminate the clipping distortion problem in DCO-OFDM and increase the spectral efficiency which is halved in ACO-OFDM and U-OFDM. In NDC-OFDM, after the DCO-OFDM modulation block, symbols are transmitted by different LEDs. The positive OFDM symbol is transmitted by one LED and the negative symbol is transmitted by another LED. As the LED can only transmit positive signals, the absolute value of the negative symbol is transmitted. Thus, unlike the conventional OSM-OFDM system, the indices of transmitters in NDC-OFDM represent the signs of the transmitted signal and the absolute value of the signal is sent as optical intensity signals. As the DC-bias and the bottom clipping do not exist in NDC-OFDM, the system can save energy. Despite the fact that NDC-OFDM uses two transmitters, the energy efficiency performs better than conventional OSM-OFDM schemes when using the same number of transmitters. Moreover, the VLC system trends to be realized by multiple low power LEDs to achieve higher transmission bit rate.

The rest of this paper is organized as follows. The system model of NDC-OFDM is described in Section \uppercase\expandafter{\romannumeral2}. Section \uppercase\expandafter{\romannumeral3} presents the performance analysis of NDC-OFDM. Section \uppercase\expandafter{\romannumeral4} shows numerical and simulation results of the performance analysis and the result of a comparison between NDC-OFDM and conventional OSM-OFDM in terms of their BER performances. Finally, Section \uppercase\expandafter{\romannumeral5} concludes this paper.

\section{System Model}

\begin{figure*}[!t]
\begin{center}
\includegraphics[width=0.9\textwidth]{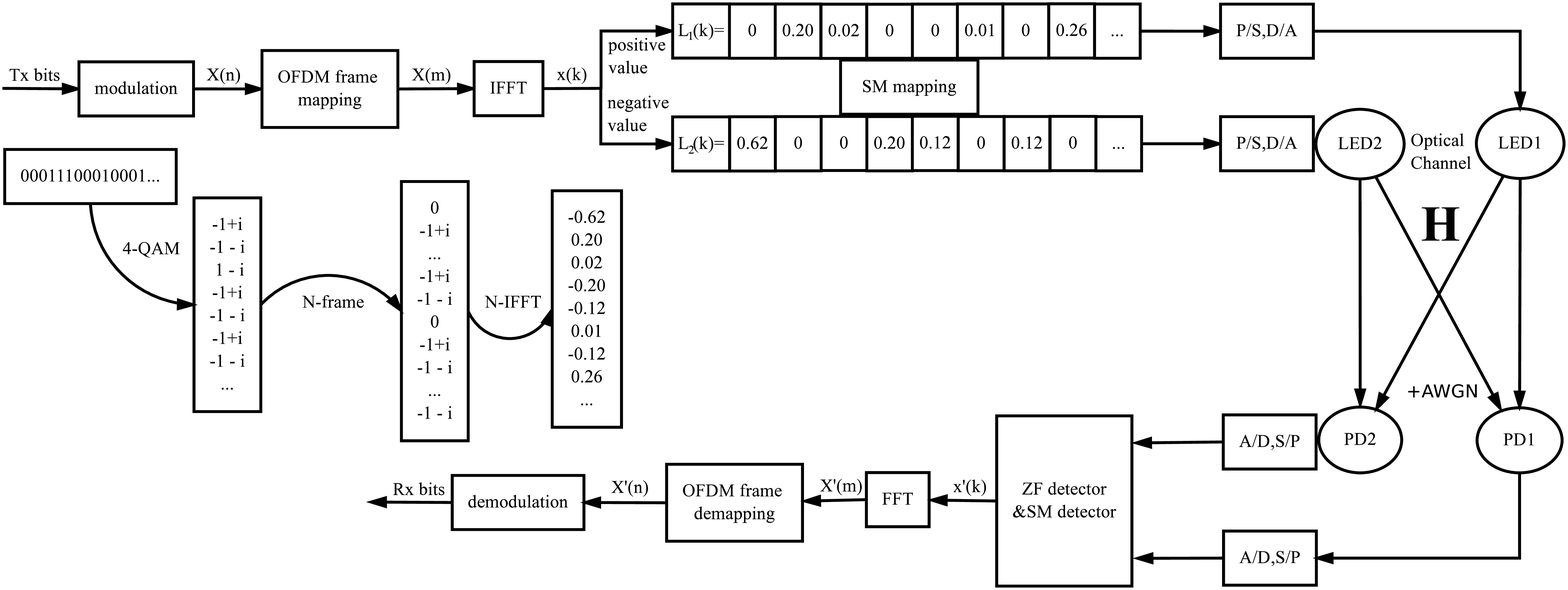}
\end{center}
\caption{Block diagram of the NDC-OFDM system}
\label{NDC-OFDM}
\end{figure*}

The NDC-OFDM system model is illustrated in Fig.~1. This system mainly aims to solve the DC-bias distortion problem in the DCO-OFDM modulation scheme. Using the indices of LEDs to transmit signs of samples ensures that transmitted samples are positive and also saves transmission energy in order to increase the spectral efficiency under a fixed power condition.

\subsection{Modulation Procedure}

At the transmitter, the input bit stream is transformed into complex symbols, $\textbf{X}(n)$, $n=0, \ \cdots, \ N/2-2$, by an $M$-QAM modulator. N is the number of OFDM subcarriers. $N/2-1$ QAM symbols are then modulated on to the first half of an OFDM frame, $\textbf{X}(m)$, $m = 0, \ \cdots, \ N-1$, and the DC subcarrier (the first subcarrier) is set to zero. Then, Hermitian symmetry is imposed on the second half of the OFDM frame. Next, the mapped subcarriers are passed through an IFFT block. Without loss of generality, the following definition of inverse discrete Fourier transform is used \cite{a0901},

\begin{equation}
\textbf{x}(k) = \frac{1}{\sqrt{N}}\sum\limits_{m = 0}^{N-1}\textbf{X}(m)\exp(\frac{j2\pi km}{N}).
\end{equation}

After the $N$-IFFT operation, the complex QAM symbols become $N$ real-valued OFDM samples, $\textbf{x}(k)$, but they are still bipolar. In the conventional DCO-OFDM system, a DC bias is added and the signal is then clipped to obtain the unipolar sample. In practice, the value of the DC bias, which is related to the average power of the OFDM symbols, is defined in \cite{as0801} as

\begin{equation}
B_{\rm DC} = \alpha\sqrt{ {\rm E}\{ \textbf{x}^2(k) \} },
\end{equation}

\noindent where $10\log_{10}(\alpha^2+1)$ is defined as the bias level in dB which depends on the constellation size. For the simple DCO-OFDM model, positive samples, which can be transmitted by LEDs, are obtained by signal clipping after a fixed power for the DC bias is added. However, the added DC biased power increases the power consumption. More importantly, if the level of DC-bias is not enough to ensure all the samples positive, the signal clipping will cause the bottom distortion problem \cite{dsh1103}.

NDC-OFDM is a novel modulation scheme which can save the transmission energy and mitigate the bottom nonlinear distortion problem. In NDC-OFDM, LEDs only send the absolute value of $\textbf{x}(k)$ and the sign of the symbol is represented by the index of the corresponding LED. According to the working principle of OSM, only one LED is activated during one symbol time. If the transmitted symbol is positive, the first LED will be activated to send the symbol. If the symbol is negative, its absolute value will be sent by the other LED. Since the absolute values of the negative samples is transmitted, this system does not need additional transmission power to obtain positive signals. Moreover, the signal clipping is not exerted in NDC-OFDM because every sample is fitted for the LED-based system.

In general, in an OFDM-based system a cyclic prefix (CP) is added to resist ISI before the samples are transmitted, but in OWC the CP is shown to have a negligible effect on the electrical SNR requirement and the spectral efficiency \cite{emh0901}. Therefore, for simplicity, it is not considered in the theoretical performance analysis in this study.

Finally, the digital signals in SM frame vectors, $\textbf{L}_1(k)$ and $\textbf{L}_2(k)$, will be transformed to analog signals and prepared to be transmitted by LEDs.

\subsection{Optical Channel}

As shown in Fig. 2, the converted optical signals will be transmitted by the corresponding LED over the optical MIMO channel $\textbf{H}$ \cite{meh1101}. Without loss of generality, a simple $N_t \times N_r$ optical channel matrix is realized,

\begin{equation} 
\textbf{H} = \left( \begin{array}{cccc} h_{11} & h_{12} & \cdots & h_{1N_t} \\ h_{21} & h_{22} & \cdots & h_{2N_t} \\ \vdots & \vdots & \ddots & \vdots \\ h_{N_r1} & h_{N_r2} & \cdots & h_{N_rN_t} \end{array} \right),
\end{equation}

\noindent where $h_{N_rN_t}$ is the channel DC gain of a directed line-of-sight (LOS) link between the receiver $N_r$ and the transmitter $N_t$. The LOS link is considered in the system model, because the multipath components are significantly weaker and can thus be neglected. The channel gain is calculated as follows \cite{kb9701},

\begin{small}
\begin{equation}
h_{\rm N_rN_t} = \left\{ \begin{array}{ll} \frac{(\beta+1)A}{2\pi d^2}\cos^\beta(\phi)T_s(\psi)g_c(\psi)\cos(\psi) ,& 0\leq\psi\leq\Psi_c \\ 0 ,& \psi>\Psi_c \end{array} \right.
\end{equation}
\end{small}

\begin{description*}
\item[Where] \hfill
\begin{description*}
\setlength{\itemsep}{0pt}
\setlength{\parskip}{0pt}
\setlength{\parsep}{0pt}
\item[$\beta$] is related to $\Phi_{1/2}$ which is the transmitter semiangle, by $\beta = -\ln 2/ \ln(\cos(\Phi_{1/2}))$,
\item[$A$] is the detector area of the PD,
\item[$d$] is the distance between the receiver $N_r$ and the transmitter $N_t$,
\item[$\phi$] is the radiant angle,
\item[$\psi$] is the incident angle,
\item[$T_s$ and $g_c$] \hspace{0.5cm} are the optical filter gain and the optical concentrator gain which depend on the properties of the receiver.
\end{description*}
\end{description*}

\subsection{Detection and Demodulation}

Through the optical MIMO channel, correlated optical signals are detected and obtained by PD receivers. The received signal can be written as

\begin{equation} 
\textbf{y} = \textbf{Hs} + \textbf{w},
\end{equation}

\noindent where $\textbf{y}$ is the $N_r$-dimensional received vector and $\textbf{s}$ is the $N_t$-dimensional transmitted signal vector. In this paper, both $N_r$ and $N_t$ are set to two. In addition, $\textbf{w}$ is the $N_r$-dimensional noise vector which is assumed to be real-valued AWGN. 

After the received optical OFDM signal is converted to an electrical signal by PD, the ZF detection is used to recover the transmitted symbols as follows \cite{fkh1201},

\begin{equation}
\textbf{g} = \textbf{H}^{-1}\textbf{y},
\end{equation}

\noindent where $\textbf{g}$ is an $N_t$-dimensional vector which contains the estimated transmitted symbols and $\textbf{H}^{-1}$ denotes the inverse of the channel matrix $\textbf{H}$. In this paper, it is assumed that the channel gain is known at the receiver. The performance of NDC-OFDM is compared with the conventional O-OFDM approach in this study and different detection methods will not effect on the comparison results. Even though the MMSE detector can also be used in the NDC-OFDM system with the known channel information and the noise coefficient, the ZF detector has been chosen as a simple and convenient detection method \cite{kkb9601}. Each element in $\textbf{g}$ represents the detected OFDM signal which has been transmitted by the corresponding LED with AWGN added at the receiver.

Before demodulating the received signal, there are two methods in which the original bipolar OFDM signal can be reconstructed. Both of the two methods are based on the principle of the modulation scheme in NDC-OFDM. The detected signal received by the first PD is set to be transmitted by the first LED which sent the positive OFDM sample. The second PD achieves the absolute value of the negative sample. In NDC-OFDM, since only one LED is activated during one time slot, only one element in $\textbf{g}$ carries the bit information and the other ones are treated as noises. According to the rules above, the first method is to subtract the negative signal block from the positive one. This is the same as the demodulation approach in ACO-OFDM. However, when reconstructed in this way, the proposed method performs 3 dB worse than bipolar OFDM for the same constellation size. This is because the subtraction of the negative block from the positive one doubles the AWGN variance for each restored bipolar OFDM signal. The other reconstruction method used in NDC-OFDM performs a significant improvement on the power efficiency which has been proved in U-OFDM \cite{tsh1201}. This method mainly aims to estimate the index of the active transmitter. The estimated index represents the sign of the transmitted information OFDM sample. The information-carried signal, afterwards, can be selected to reconstruct the bipolar OFDM signal. Compared with the former approach, this method will not double the AWGN variance for each estimated OFDM symbol. In particular, to estimate the indices of the active transmitters, the SM detector compares the values of the elements in $\textbf{g}$ as follows,

\begin{equation}
\tilde{\textbf{l}}(k) = \arg\max_{i}(\textbf{G}(i,k)) , i = 1,\cdots,N_t,
\end{equation}

\noindent where $\textbf{G}$ is the $N_t \times N$ equalized matrix which contains all the estimated transmitted symbols and $\tilde{\textbf{l}}$ is an $N$-dimensional vector which contains all the estimated indices. As noted, there are two transmitters and two receivers. If $\tilde{\textbf{l}}(k)$ is equal to one, this means that the symbol received at the time instant $k$ is transmitted from the first LED. Therefore this symbol is a positive-valued OFDM symbol. If $\tilde{\textbf{l}}(k)$ is two, a negative symbol is transmitted by LED2. As a consequence, the estimated OFDM symbols sequence is

\begin{equation}
\textbf{x}'(k) = \left\{ \begin{array}{ll} \textbf{G}(\tilde{\textbf{l}}(k),k), & \tilde{\textbf{l}}(k) = 1, \\ -\textbf{G}(\tilde{\textbf{l}}(k),k), & \tilde{\textbf{l}}(k) = 2. \end{array} \right.
\end{equation}

\noindent In an ideal scenario, if there is no AWGN, $\textbf{x}'(k)$ should be the same as $\textbf{x}(k)$. In this paper, the sign-selected estimation has been chosen. After recovering the OFDM symbols, $\textbf{x}'(k)$ is passed through the conventional OFDM demodulation block to obtain received QAM symbols,

\begin{equation}
\textbf{X}'(m) = \frac{1}{\sqrt{N}}\sum\limits_{k = 0}^{N-1}\textbf{x}'(k)\exp(\frac{-j2\pi km}{N}).
\end{equation}

The $N/2-1$ data-carrying symbols in $\textbf{X}'(m)$ can be extracted and then demodulated using a maximum likelihood (ML) detector in order to obtain the output bit stream.

\section{Performance Analysis}

The theoretical BER performance of NDC-OFDM is computed and presented in this section in order to demonstrate the correctness of experimental results. Moreover, the spectral efficiency of NDC-OFDM is analysed and compared with the spectral efficiency of other OFDM methods in the OSM system.

\subsection{Theoretical Performance of NDC-OFDM}

To calculate the theoretical BER of NDC-OFDM, the following mathematical notations and formulas should be defined. In this paper, $\sigma_{\rm n}$ is the standard deviation of the AWGN, i.e., $\sigma_{\rm n} = \sqrt{N_0/2}$, where $N_0/2$ is the variance of the AWGN.The constant, $\sigma_{\rm s}$, is the standard deviation of the real OFDM signals which have been modulated and are ready to be transmitted by LEDs. For the analytical calculation, $\sigma_{\rm s}$ is defined as follow,

\begin{equation}
\sigma_{\rm s} = \sqrt{E_{\rm b} \log_2(M) \frac{N-2}{2NN_t}},
\end{equation}

\noindent where $E_{\rm b}$ is the electrical energy per bit. $E_{\rm b}/N_0$ is the metric of the BER performance. $\phi(x)$ is the standard normal distribution probability density function, i.e.,

\begin{equation}
\phi(x) = \frac{1}{\sqrt{2\pi}} e^{-\frac{x^2}{2}}
\end{equation}

\noindent $\textit{1}(x)$ is the step function, i.e.,

\begin{equation}
\textit{1}(x) = \left\{
\begin{array}{ll}
1 & {\rm if} \ x > 0 \\
0 & {\rm if} \ x\leq 0
\end{array}
\right.
\end{equation}

\noindent ${\rm sgn}(s)$ is the sign function, i.e.,

\begin{equation}
{\rm sgn}(s) = \left\{
\begin{array}{lll}
-1 & {\rm if} \ s < 0 \\
0 & {\rm if} \ s = 0 \\
1 & {\rm if} \ s > 0.
\end{array}
\right.
\end{equation}

In following theoretical expressions, a $2 \times 2$ MIMO channel is considered,

\begin{equation}
\textbf{H} = \left( \begin{array}{cc} h_{11} & h_{12} \\ h_{21} & h_{22} \end{array} \right).
\end{equation}

\noindent Since the attenuation gain of the channel has a limited effect on the results of the analytical BER performance, for simplicity, the coefficients in $\textbf{H}$ are normalized to one and they simply represent the correlation coefficients of the channel. The ZF detection needs to use the inverse matrix of the channel matrix, $\textbf{H}^{-1}$, to eliminate the channel effect on information samples. The inverse matrix is represented by $\textbf{C}$, i.e.,

\begin{equation}
\textbf{C} = \left( \begin{array}{cc} c_{11} & c_{12} \\ c_{21} & c_{22} \end{array} \right).
\end{equation}

\begin{figure*}[ht]
\begin{equation}
Pr_{\rm c}(s, n_1, n_2) = \left\{
\begin{array}{ll}
\frac{1}{\sigma_{\rm n}^2}\phi(\frac{n_1}{\sigma_{\rm n}})\phi(\frac{n_2}{\sigma_{\rm n}}){\textit{1}}(|s|+(c_{11}-c_{21})n_1+(c_{12}-c_{22})n_2), & s \geqslant 0,
\\ \frac{1}{\sigma_{\rm n}^2}\phi(\frac{n_1}{\sigma_{\rm n}})\phi(\frac{n_2}{\sigma_{\rm n}}){\textit{1}}(|s|+(c_{21}-c_{11})n_1+(c_{22}-c_{12})n_2), & s < 0.
\end{array}
\right.
\label{Prc}
\end{equation}
\begin{equation}
Pr_{\rm w}(s, n_1, n_2) = \left\{
\begin{array}{ll}
\frac{1}{\sigma_{\rm n}^2}\phi(\frac{n_1}{\sigma_{\rm n}})\phi(\frac{n_2}{\sigma_{\rm n}}){\textit{1}}(-|s|-(c_{11}-c_{21})n_1-(c_{12}-c_{22})n_2), & s \geqslant 0,
\\ \frac{1}{\sigma_{\rm n}^2}\phi(\frac{n_1}{\sigma_{\rm n}})\phi(\frac{n_2}{\sigma_{\rm n}}){\textit{1}}(-|s|-(c_{21}-c_{11})n_1-(c_{22}-c_{12})n_2), & s < 0.
\end{array}
\right.
\label{Prw}
\end{equation}
\hrulefill
\end{figure*}

Based on the theoretical analysis method of the nonlinear transmission in \cite{tsh1201}, the analysis in this study mainly aims to calculate the probability of the correct and incorrect detection in which the effects of the ZF detection and the nonlinear OFDM demodulation should be taken into consideration. In NDC-OFDM, two receivers obtain optical OFDM samples over the MIMO channel at the same time. After the ZF detection, the unipolar OFDM symbols are recovered with the enhanced AWGN. Symbols detected by the first PD come from the first LED which are originally positive symbols. Symbols detected by the second PD are transmitted by the second LED which are the absolute values of the negative symbols. If there is no noise in the system, the received symbols should be the same as the transmitted symbols. In the theoretical analysis model, the AWGNs are considered as two independent random variables, $n_1$ and $n_2$, which follow the standard normal distribution with the standard deviation, $\sigma_{\rm n}$. Since the ZF detection is used in the system, the noise is enhanced after removing the channel crosstalk. Most importantly, the AWGN in one receiver has an impact on the variance of the noise in the other receiver. Considering this condition, the correctly detected probability for the identical symbol is presented in (\ref{Prc}). This depends on a random value of $n_1$, a random value of $n_2$, the inverse matrix of the channel and the original bipolar symbol, $s$. The bipolar OFDM symbols also follow the Gaussian distribution. Likewise, the incorrectly detected probability is given in (\ref{Prw}). With the identical $n_1$, $n_2$ and $s$, the correctly detected OFDM sample is expressed as follow,

\begin{equation}
x_{\rm c} = \left\{
\begin{array}{ll}
|s|+c_{11}n_1+c_{12}n_2, & s \geqslant 0,
\\ |s|+c_{21}n_1+c_{22}n_2, & s < 0.
\end{array}
\right.
\end{equation}

\noindent For all possible values of $n_1$ and $n_2$ and the identical OFDM sample, the mean of $x_{\rm c}$ is,

\begin{equation}
f_{\rm c}(s) = 
\frac{{\rm sgn}(s)\int_{-\infty}^{\infty}\int_{-\infty}^{\infty}x_{\rm c}Pr_{\rm c}(s, n_1, n_2)\,{\rm d}n_1{\rm d}n_2}
{\int_{-\infty}^{\infty}\int_{-\infty}^{\infty}Pr_{\rm c}(s, n_1, n_2)\,{\rm d}n_1{\rm d}n_2}.
\end{equation}

\noindent Moreover, the variance of the correctly detected sample has the following value,

\begin{equation}
v_{\rm c}(s) = 
\frac{\int_{-\infty}^{\infty}\int_{-\infty}^{\infty}x_{\rm c}^2 Pr_{\rm c}(s, n_1, n_2)\,{\rm d}n_1{\rm d}n_2}
{\int_{-\infty}^{\infty}\int_{-\infty}^{\infty}Pr_{\rm c}(s, n_1, n_2)\,{\rm d}n_1{\rm d}n_2} - f_{\rm c}^2(s).
\end{equation}

\noindent Based on the Central Limit Theorem (CLT), after the FFT is exposed in the OFDM demodulation process, the variance, $v_{\rm c}(s)$ will be a part of the variance of the AWGN in the frequency domain. For the detection of NDC-OFDM, the incorrect determination will enhance the variance of the AWGN, but this enhancement is much less than the detection method of the conventional ACO-OFDM which doubles the variance. For incorrect detection, the selected OFDM sample is calculated as,

\begin{equation}
x_{\rm w} = \left\{
\begin{array}{ll}
c_{21}n_1+c_{22}n_2, & s \geqslant 0,
\\ c_{11}n_1+c_{12}n_2, & s < 0.
\end{array}
\right.
\end{equation}

\noindent The mean and the variance of this value have the following forms,

\begin{equation}
f_{\rm w}(s) = 
\frac{{\rm -sgn}(s)\int_{-\infty}^{\infty}\int_{-\infty}^{\infty}x_{\rm w}Pr_{\rm w}(s, n_1, n_2)\,{\rm d}n_1{\rm d}n_2}
{\int_{-\infty}^{\infty}\int_{-\infty}^{\infty}Pr_{\rm w}(s, n_1, n_2)\,{\rm d}n_1{\rm d}n_2},
\end{equation}

\begin{equation}
v_{\rm w}(s) = 
\frac{\int_{-\infty}^{\infty}\int_{-\infty}^{\infty}x_{\rm w}^2 Pr_{\rm w}(s, n_1, n_2)\,{\rm d}n_1{\rm d}n_2}
{\int_{-\infty}^{\infty}\int_{-\infty}^{\infty}Pr_{\rm w}(s, n_1, n_2)\,{\rm d}n_1{\rm d}n_2} - f_{\rm w}^2(s).
\end{equation}

\noindent Since the OFDM samples, $s$, follow a Gaussian distribution, for all possibility of $s$, the average variances of the correct and incorrect detections are,

\begin{equation}
\bar{v}_{\rm c} = \int_{-\infty}^{\infty}v_{\rm c}(s)\frac{1}{\sigma_{\rm s}}\phi(\frac{s}{\sigma_{\rm s}}){\rm d}s,
\end{equation}

\noindent and

\begin{equation}
\bar{v}_{\rm w} = \int_{-\infty}^{\infty}v_{\rm w}(s)\frac{1}{\sigma_{\rm s}}\phi(\frac{s}{\sigma_{\rm s}}){\rm d}s.
\end{equation}

\noindent These variances will constitute the variance of the AWGN in frequency domain.

After the NDC-OFDM detection, the selected symbols should be demodulated to QAM symbols by FFT. For NDC-OFDM, the demodulation procedure is treated as a nonlinear transformation. According to the Bussgang theorem \cite{tech:jbuss01}, if an independent Gaussian random variable, $X$, passes through a nonlinear transformation, $f(X)$, which has the following properties,

\begin{equation}
\left\{
\begin{array}{lll}
f(X) = \alpha_{\rm d}X + Y \\
{\mathcal E}[XY] = 0 \\
\alpha_{\rm d} = {\rm const},
\end{array}
\right.
\label{BT}
\end{equation}

\noindent where ${\mathcal E}[.]$ expresses the statistical expectation. Using the properties above, the nonlinear distortion in an OFDM-based system can be equivalent to a gain factor, $\alpha_{\rm d}$, and an additional noise, $Y$ \cite{tsh1201}. In NDC-OFDM, $X$ is equal to the value of the transmitted symbol, $s$, and $Y$ is a noise component which is a Gaussian random variable non-correlated with $X$. After exposing FFT, the variance of $Y$ will be composed of the other part of the variance of the AWGN in the frequency domain and $\alpha_{\rm d}$ will enhance the mean value of the information-carried symbol in each modulated subcarrier. According to \ref{BT}, $\alpha_{\rm d}$ can be derived as,

\begin{equation}
\alpha_{\rm d} = \frac{{\mathcal E}[Xf(X)]}{\sigma_{\rm X}^2}
\label{alpha}
\end{equation}

\noindent where $\sigma_{\rm X}$ is the standard deviation of $X$, which is equal to $\sigma_{\rm s}$ in this study. Since the additional noise, $Y$, follows a Gaussian distribution with a zero mean, the variance of $Y$ can be calculated as,


\begin{equation}
\begin{aligned}
\sigma_{\rm Y}^2 & = {\mathcal E}[Y^2] - {\mathcal E}[Y]^2 = {\mathcal E}[Y^2] \\
& = {\mathcal E}[(f(X) - \alpha_{\rm d}X)^2] \\
& = {\mathcal E}[f^2(X)] - \alpha_{\rm d}^2\sigma_{\rm X}^2.
\end{aligned}
\label{sigmaY}
\end{equation}

\noindent From (\ref{alpha}) and (\ref{sigmaY}), the values of the constant and the variance for the correct detection are expressed as,

\begin{equation}
\alpha_{\rm c} = \frac {\int_{-\infty}^{\infty}sf_{\rm c}(s)\frac{1}{\sigma_{\rm s}}\phi(\frac{s}{\sigma_{\rm s}}){\rm d}s} {\sigma_{\rm s}^2},
\end{equation}

\begin{equation}
y_{\rm c} = \int_{-\infty}^{\infty}f_{\rm c}^2(s)\frac{1}{\sigma_{\rm s}}\phi(\frac{s}{\sigma_{\rm s}}){\rm d}s - \alpha_{\rm c}^2\sigma_{\rm s}^2,
\end{equation}

\noindent where $y_{\rm c}$ is the variance. For the incorrect detection, the constant, $\alpha_{\rm w}$, and the variance, $y_{\rm w}$, are calculated as,

\begin{equation}
\alpha_{\rm w} = \frac {\int_{-\infty}^{\infty}sf_{\rm w}(s)\frac{1}{\sigma_{\rm s}}\phi(\frac{s}{\sigma_{\rm s}}){\rm d}s} {\sigma_{\rm s}^2},
\end{equation}

\begin{equation}
y_{\rm w} = \int_{-\infty}^{\infty}f_{\rm w}^2(s)\frac{1}{\sigma_{\rm s}}\phi(\frac{s}{\sigma_{\rm s}}){\rm d}s - \alpha_{\rm w}^2\sigma_{\rm s}^2.
\end{equation}

From (\ref{Prc}), the probability of the correct detection during an active time slot is,

\begin{equation}
d_{\rm c} = \int_{-\infty}^{\infty} \int_{-\infty}^{\infty} \int_{-\infty}^{\infty} \frac{1}{\sigma_{\rm s}}\phi(\frac{s}{\sigma_{\rm s}})Pr_{\rm c}(s, n_1, n_2){\rm d}n_1{\rm d}n_2{\rm d}s
\end{equation}

\noindent For a large number of samples in a NDC-OFDM frame, the number of correctly and incorrectly detected active samples have a ratio which corresponds to the probabilities for correct and incorrect detection. According to (\ref{BT}), the nonlinear transmission will add a gain factor to the sample. For the theoretical analysis, the average value of the gain factor will enhance the average energy of the transmitted bits. The average gain factor is calculated as,

\begin{equation}
\bar{\alpha} = d_{\rm c}\alpha_{\rm c} + (1 - d_{\rm c})\alpha_{\rm w}
\end{equation}

\noindent As noted above, the variance of the detection, $\bar{v}_{\rm c}$ and $\bar{v}_{\rm w}$, and the variance of the nonlinear transmission, $y_{\rm c}$ and $y_{\rm w}$ will constitute the average noise variance of the system in the frequency domain, i.~e.,

\begin{equation}
\bar{N} = d_{\rm c}(\bar{v}_{\rm c} + y_{\rm c}) + (1-d_{\rm c})(\bar{v}_{\rm w} + y_{\rm w})
\end{equation}

\noindent Thus, the average ${\rm SNR_{elec}}$ per bit can be achieved from the known value of $E_{\rm b,elec}$ and the calculated values of $\bar{\alpha}$ and $\bar{N}$ as,

\begin{equation}
{\rm SNR_{elec}} = \frac{\bar{\alpha}^2E_{\rm b,elec}}{\bar{N}}
\end{equation}

Using the analytical expression for the BER performance of $M-$QAM O-OFDM in \cite{dsh1102}, the theoretical BER performance of NDC-OFDM can be calculated as,

\begin{equation}
\begin{aligned}
{\rm BER_{NDC}} = \frac{4(\sqrt{M}-1)}{\sqrt{M}\log_2(M)}Q\left(\sqrt{\frac{3\log_2(M)}{M-1}{\rm SNR_{elec}}}\right)
\\ +\frac{4(\sqrt{M}-2)}{\sqrt{M}\log_2(M)}Q\left(3\sqrt{\frac{3\log_2(M)}{M-1}{\rm SNR_{elec}}}\right)
\end{aligned}
\end{equation}

\subsection{Spectral Efficiency Comparison}

NDC-OFDM is realized in the OSM system which can also apply ACO-OFDM and DCO-OFDM as the modulation schemes. For fair comparisons, NDC-OFDM, ACO-OFDM and DCO-OFDM are built with an OSM system. For NDC-OFDM, the indices of LEDs is used to carry the sign information. For ACO-OFDM and DCO-OFDM, the indices carry additional information bits according to the conventional principle of the OSM system. For a normal OSM system with the simple $M-$QAM, the spectral efficiency is calculated by considering both the transmitted signal information bits and the indices-carried bits, i.e., $R_{\rm OSM} = \log_2(MN_t) \ {\rm bits/s/Hz}$ \cite{meh1101}. For NDC-OFDM, since the Hermitian symmetry of O-OFDM decreases the spectral efficiency by half and there is no information bit carried by the indices, therefore the spectral efficiency of NDC-OFDM is defined as,

\begin{equation}
R_{\rm NDC-OFDM} = \frac{N-2}{2N} \left[ \log_2(M_1N_t)-1 \right] {\rm{bits/s/Hz}},
\label{NDCSE}
\end{equation}

\noindent Since in NDC-OFDM two different signs of the samples should be represented respectively, the number of the LEDs, $N_t$, should be even. For DCO-OFDM in the OSM system, the spectral efficiency is halved by the Hermitian symmetry, so it is expressed as,

\begin{equation}
R_{\rm DCO-OFDM} = \frac{N-2}{2N}\log_2(M_2N_t) \rm{bits/s/Hz}.
\label{DCOSE}
\end{equation}

\noindent In ACO-OFDM, because only half of the subcarriers are modulated, the spectral efficiency should have an additional 50$\%$ reduction. In the OSM system, the actual spectral efficiency of ACO-OFDM is,

\begin{equation}
R_{\rm ACO-OFDM} = \frac{1}{4}\log_2(M_3N_t) \rm{bits/s/Hz}.
\label{ACOSE}
\end{equation}

\noindent In \ref{NDCSE}, \ref{DCOSE} and \ref{ACOSE}, $M_1$, $M_2$ and $M_3$ denote the constellation size of QAM in the three O-OFDM modulation schemes respectively. In this study, the size of the OFDM frame, $N$, is set to 2048, so the coefficient, $\frac{N-2}{2N}$, in (\ref{NDCSE}) and (\ref{DCOSE}) can be treated as 1/2. When NDC-OFDM, DCO-OFDM and ACO-OFDM in the OSM system have the same spectral efficiencies,i.e., $R_{\rm NDC-OFDM}=R_{\rm DCO-OFDM}=R_{\rm ACO-OFDM}$, the constellation sizes of these three methods have the following relationship,

\begin{equation}
M_1 = 2M_2 = \sqrt{2M_3}.
\label{RelatM}
\end{equation}

\begin{figure}[!t]
\begin{center}
\includegraphics[width=0.48\textwidth]{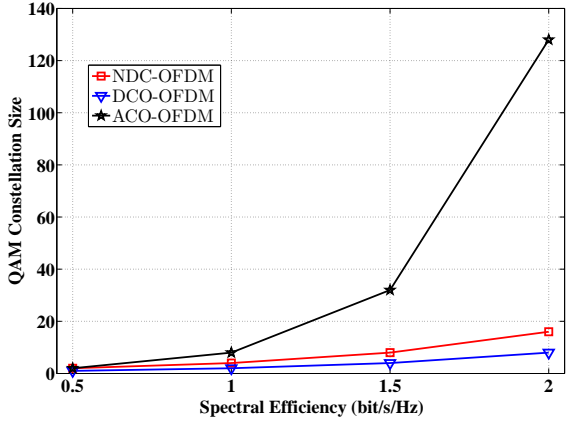}
\caption{Constellation size vs. spectral efficiency for NDC-OFDM, DCO-OFDM and ACO-OFDM in the OSM system}
\end{center}
\label{SECom}
\end{figure}

Fig.~2 shows that constellation sizes are used to reach the same spectral efficiencies between 0.5~bits/s/Hz and 2~bits/s/Hz for NDC-OFDM, DCO-OFDM and ACO-OFDM in the OSM system. It can be seen that NDC-OFDM needs a little higher constellation size to reach the same spectral efficiency as DCO-OFDM. Most importantly, if the spectral efficiency of ACO-OFDM is equal to the spectral efficiency of NDC-OFDM, the constellation size will increase exponentially which costs the system complexity. For the high spectral efficiencies, such as between 3.5~bits/s/Hz and 5.5~bits/s/Hz, the increase of the constellation size of ACO-OFDM becomes very large as shown in Table~\ref{SE}. Although the constellation size can be as large as required to achieve the higher spectral efficiency, the data in the table for ACO-OFDM is unrealistic and unattainable. For the high speed optical transmission, NDC-OFDM and DCO-OFDM are more realistic.

\begin{table}[!t]
\renewcommand{\arraystretch}{1.5}
\caption{Constellation Sizes Comparison}
\centering
\begin{tabular}{|c|*{3}{c|}}\hline 
\backslashbox{SE (bits/s/Hz)} {Method} & NDC & DCO & ACO\\\hline 
3.5 & 128  & 64   & 8192   \\\hline 
4   & 256  & 128  & 32768  \\\hline 
4.5 & 512  & 256  & 131072 \\\hline 
5   & 1024 & 512  & 524288 \\\hline 
5.5 & 2048 & 1024 & 2097152\\\hline 
\end{tabular} 
\label{SE}
\end{table}

\section{Numerical and Simulation Results}

\subsection{Analytical Results}

\begin{figure}[!t]
\begin{center}
\includegraphics[width=0.48\textwidth]{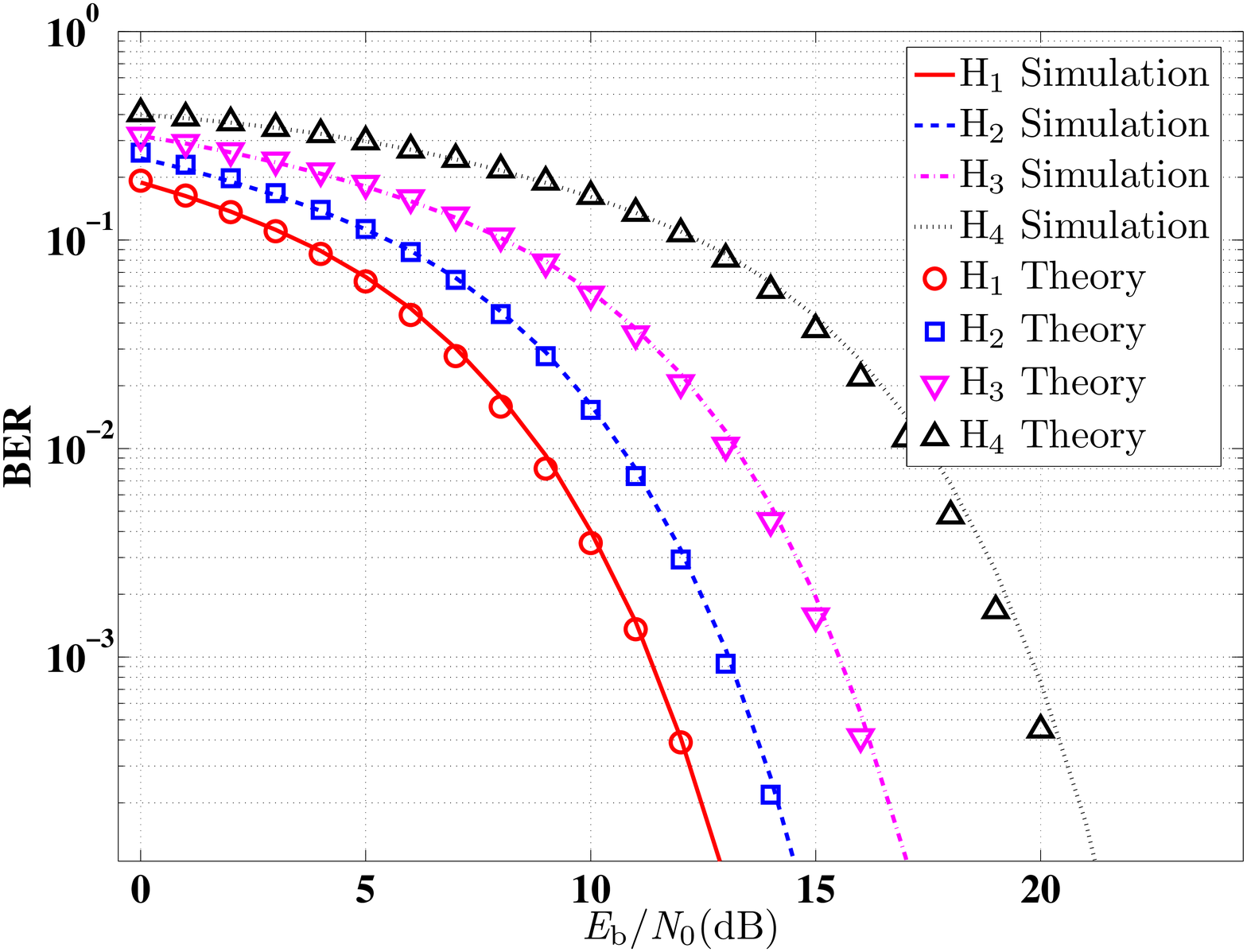}
\caption{Comparison between analytical and simulation results for symmetrical ideal channels: ${\textbf H_1}, {\textbf H_2}, {\textbf H_3}, {\textbf H_4}$}
\end{center}
\label{SymIdealChannel}
\end{figure}

\begin{figure}[!t]
\begin{center}
\includegraphics[width=0.48\textwidth]{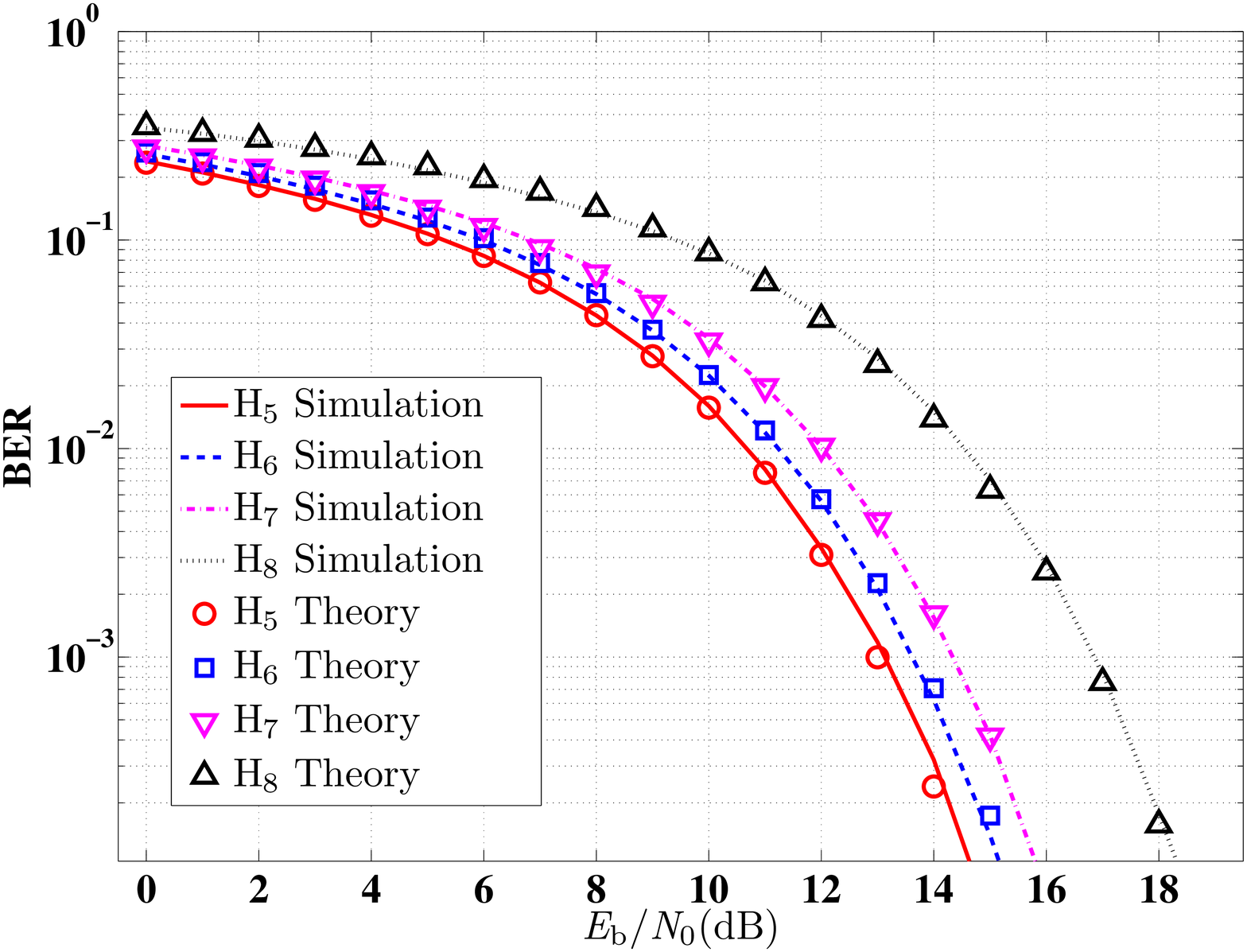}
\caption{Comparison between analytical and simulation results for asymmetrical ideal channels: ${\textbf H_5}, {\textbf H_6}, {\textbf H_7}, {\textbf H_8}$}
\end{center}
\label{AsymIdealChannel}
\end{figure}

As noted in Section~\uppercase\expandafter{\romannumeral3}, simple correlated channels are considered to test the correctness of the theoretical analysis. Symmetrical ideal channels are set as follows,

\begin{equation}
\begin{array}{ll}
{\textbf H_1} = \left( \begin{array}{cc} 1 & 0 \\ 0 & 1 \end{array} \right), &
{\textbf H_2} = \left( \begin{array}{cc} 1 & 0.3 \\ 0.3 & 1 \end{array} \right), \\ \\
{\textbf H_3} = \left( \begin{array}{cc} 1 & 0.5 \\ 0.5 & 1 \end{array} \right), &
{\textbf H_4} = \left( \begin{array}{cc} 1 & 0.7 \\ 0.7 & 1 \end{array} \right).
\end{array}
\label{SymChannel}
\end{equation}

\noindent The coefficients in (\ref{SymChannel}) reflect the correlation of optical channels. A high value of coefficient indicates a high level of correlation. Without loss of generality, asymmetrical ideal channels are also tested in this study,

\begin{equation}
\begin{array}{ll}
{\textbf H_5} = \left( \begin{array}{cc} 1 & 0 \\ 0 & 0.7 \end{array} \right), &
{\textbf H_6} = \left( \begin{array}{cc} 1 & 0 \\ 0.3 & 0.7 \end{array} \right), \\ \\
{\textbf H_7} = \left( \begin{array}{cc} 1 & 0.5 \\ 0 & 0.7 \end{array} \right), &
{\textbf H_8} = \left( \begin{array}{cc} 1 & 0.5 \\ 0.3 & 0.7 \end{array} \right).
\end{array}
\end{equation}

\noindent Since the constellation size can be treated as a factor which will not effect on the result of the integration in the theoretical expressions, 16-QAM is chosen in the implementation and for simplicity, the variance of the noise, $\sigma_n$, is set to $\sqrt{0.01}$.

Fig.~3 shows the comparison between analytical and simulation results for the symmetrical ideal channels. The performance of the theoretical model for the asymmetrical ideal channels is compared with Monte Carlo simulations in Fig.~4. It can be seen that the analytical model and the simulations show close agreement.

\subsection{NDC-OFDM, ACO-OFDM and DCO-OFDM Performance Comparison}

\begin{figure}[!t]
\begin{center}
\includegraphics[width=0.48\textwidth]{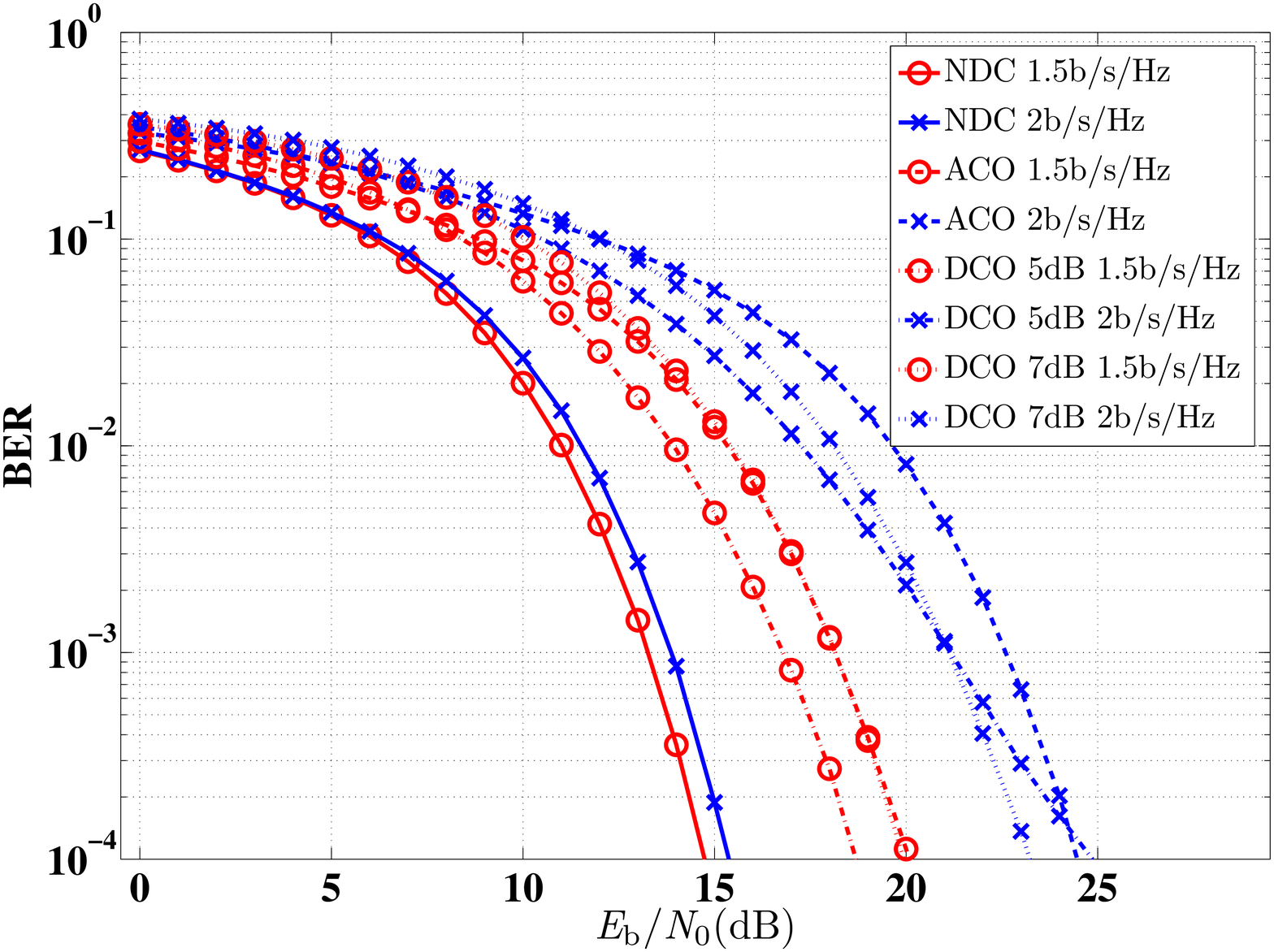}
\caption{NDC-OFDM, ACO-OFDM and DCO-OFDM Performance Comparison over ${\textbf H_{\rm Prac_1}}$}
\end{center}
\label{Hprac1}
\end{figure}

\begin{figure}[!t]
\begin{center}
\includegraphics[width=0.48\textwidth]{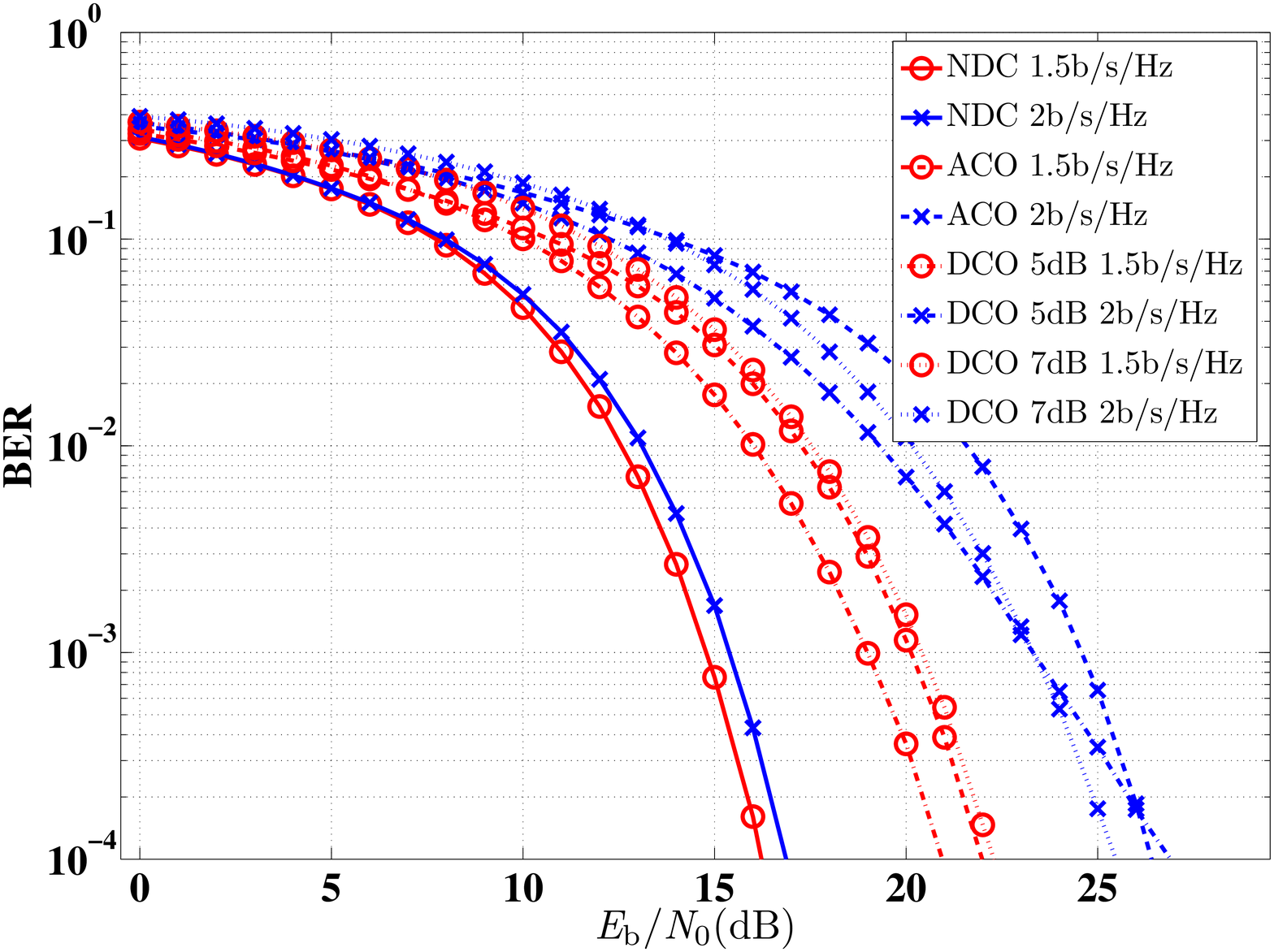}
\caption{NDC-OFDM, ACO-OFDM and DCO-OFDM Performance Comparison over ${\textbf H_{\rm Prac_2}}$}
\end{center}
\label{Hprac2}
\end{figure}

The Monte Carlo simulation results for NDC-OFDM, ACO-OFDM and DCO-OFDM are shown in this section. The BER performance of NDC-OFDM is compared with ACO-OFDM and DCO-OFDM over different practical optical MIMO channels which are chosen from \cite{fkh1201}. In \cite{fkh1201}, a generic $4\times4$ indoor scenario is considered with intensity modulated optical wireless links with LOS characteristics. In this paper, despite the fact that MIMO channels with a larger size can be realized in NDC-OFDM and OSM-OFDM, $2\times2$ optical MIMO channels are considered to test properties and advantages simply and easily. Thus, $2\times2$ optical MIMO links are extracted from original $4\times4$ optical channels, taking into account the symmetrical and asymmetrical cases,

\begin{equation}
\begin{array}{l}
{\textbf H_{\rm Prac_1}} = 10^{-5} \times \left( \begin{array}{cc} 0.1889 & 0.0713 \\ 0.0713 & 0.1889 \end{array} \right), \\ \\
{\textbf H_{\rm Prac_2}} = 10^{-5} \times \left( \begin{array}{cc} 0.3847 & 0.1889 \\ 0.1889 & 0.3847 \end{array} \right), \\ \\
{\textbf H_{\rm Prac_3}} = 10^{-5} \times \left( \begin{array}{cc} 0.1889 & 0.0713 \\ 0.1157 & 0.1889 \end{array} \right), \\ \\
{\textbf H_{\rm Prac_4}} = 10^{-5} \times \left( \begin{array}{cc} 0.3847 & 0.2691 \\ 0.1889 & 0.3847 \end{array} \right).
\end{array}
\end{equation}

\noindent ${\textbf H_{\rm Prac_1}}$, ${\textbf H_{\rm Prac_2}}$, ${\textbf H_{\rm Prac_3}}$ and ${\textbf H_{\rm Prac_4}}$ represent simple practical optical MIMO channels in the indoor scenario. Without loss of fairness, the spectral efficiency of these three methods should be same in order to compare power efficiencies. In the comparison, spectral efficiencies are set to 1.5 b/s/Hz and 2 b/s/Hz. According to (\ref{RelatM}), 8-QAM and 16-QAM are thus chosen in the simulation of NDC-OFDM; these are double than the constellation size of DCO-OFDM; and for ACO-OFDM, the modulation orders are 32 and 128. As noted in Section~\uppercase\expandafter{\romannumeral2}, a fixed level of DC-bias needs to be added in DCO-OFDM. The lower level might cause the nonlinear distortion and the higher level would be energy inefficient. To show these two cases and to simulate a real situation, 5~dB and 7~dB DC-bias are chosen in the implementation.

Fig.~5 and Fig.~6 show the performance of NDC-OFDM, ACO-OFDM and DCO-OFDM with OSM over the symmetrical optical MIMO channels, ${\textbf H_{\rm Prac_1}}$ and ${\textbf H_{\rm Prac_2}}$. It shows that when the spectral efficiency is 1.5 b/s/Hz, NDC-OFDM has around 3.5~dB power efficiency better than the 5~dB DCO-OFDM and nearly 5~dB better than ACO-OFDM and the 7~dB DCO-OFDM. In this case, there is no nonlinear distortion in DCO-OFDM. However, when the spectral efficiency is 2 b/s/Hz, the nonlinear distortion appears in the BER performance of the 5~dB DCO-OFDM. Moreover, at that time, NDC-OFDM can save 7~dB energy compared with DCO-OFDM. It means that with the increase of the spectral efficiency, the performance of NDC-OFDM is closed to the unipolar line and this is shown in \cite{tsh1201}. ${\textbf H_{\rm Prac_2}}$ is also a symmetrical channel but the correlation is different with ${\textbf H_{\rm Prac_1}}$. Fig.~6 indicates the performance of the three methods over ${\textbf H_{\rm Prac_2}}$. It seems that the change of the correlation of the channel may not change the relationship between the performance of each method. NDC-OFDM is also the most power-saving method in this case.

The performance of NDC-OFDM, ACO-OFDM and DCO-OFDM over the asymmetrical channels is indicated in Fig.~7 and Fig.~8. As shown in Fig.~7, for the 1.5~b/s/Hz spectral efficiency, the NDC-OFDM can save 5~dB energy more than 5~dB DCO-OFDM and around 7~dB more than 7~dB DCO-OFDM and ACO-OFDM in ${\textbf H_{\rm Prac_3}}$. In the 2~b/s/Hz case, when BER = $10^{-4}$, there is a 9~dB improvement between NDC-OFDM and 7~dB DCO-OFDM. NDC-OFDM gives better power efficiency than the other two methods. In the higher correlation channel, ${\textbf H_{\rm Prac_4}}$, the improvement increases to 10~dB (Fig.~8). Moreover, when the highest correlation channel, ${\textbf H_{\rm Prac_4}}$, is considered, NDC-OFDM has the greater superiority over DCO-OFDM and ACO-OFDM. This means that the transmission method of NDC-OFDM performs the ability of the anti-crosstalk better than the conventional O-OFDM methods in the OSM system.

\begin{figure}[!t]
\begin{center}
\includegraphics[width=0.48\textwidth]{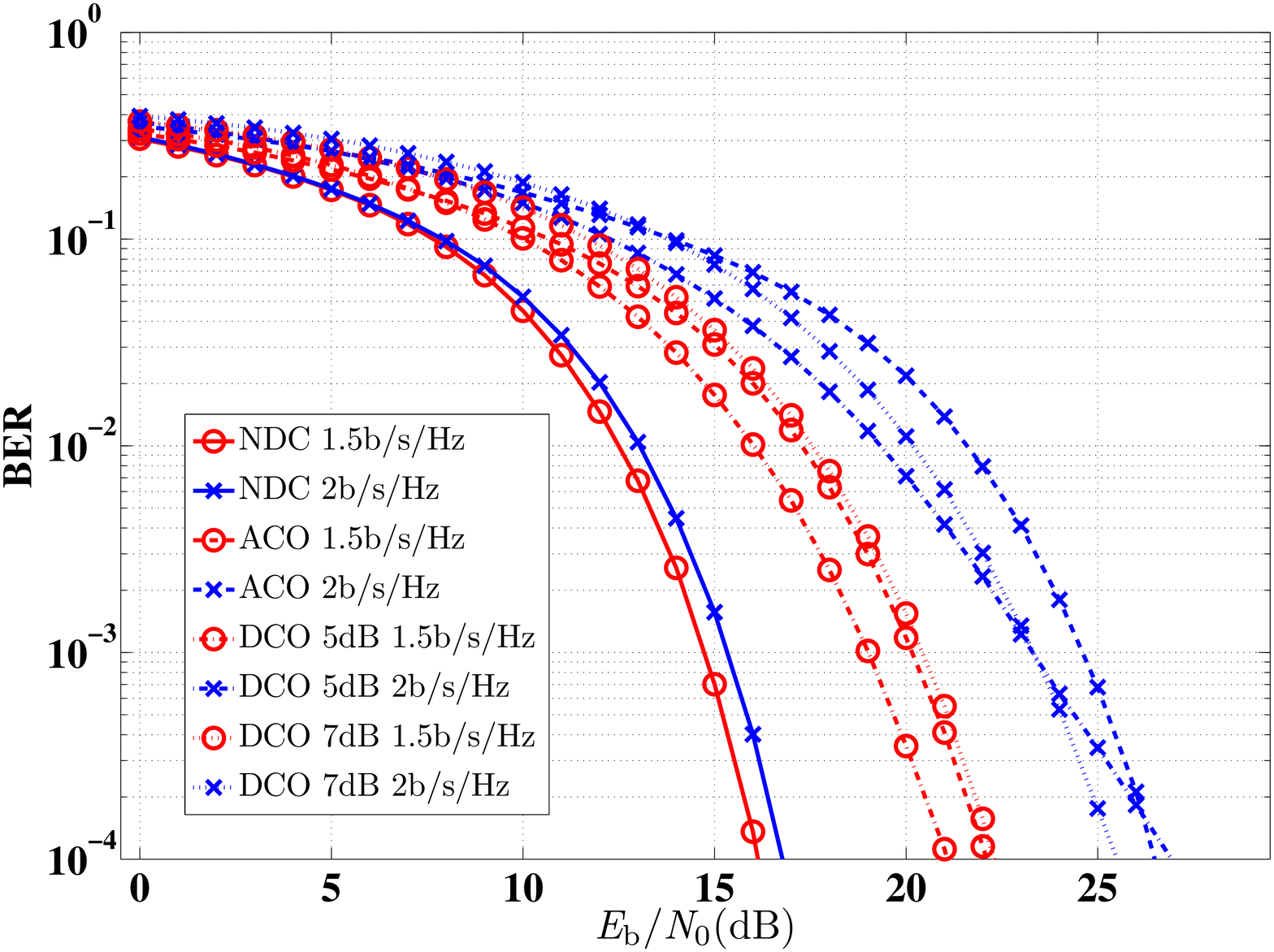}
\caption{NDC-OFDM, ACO-OFDM and DCO-OFDM Performance Comparison over ${\textbf H_{\rm Prac_3}}$}
\end{center}
\label{Hprac3}
\end{figure}

\begin{figure}[!t]
\begin{center}
\includegraphics[width=0.48\textwidth]{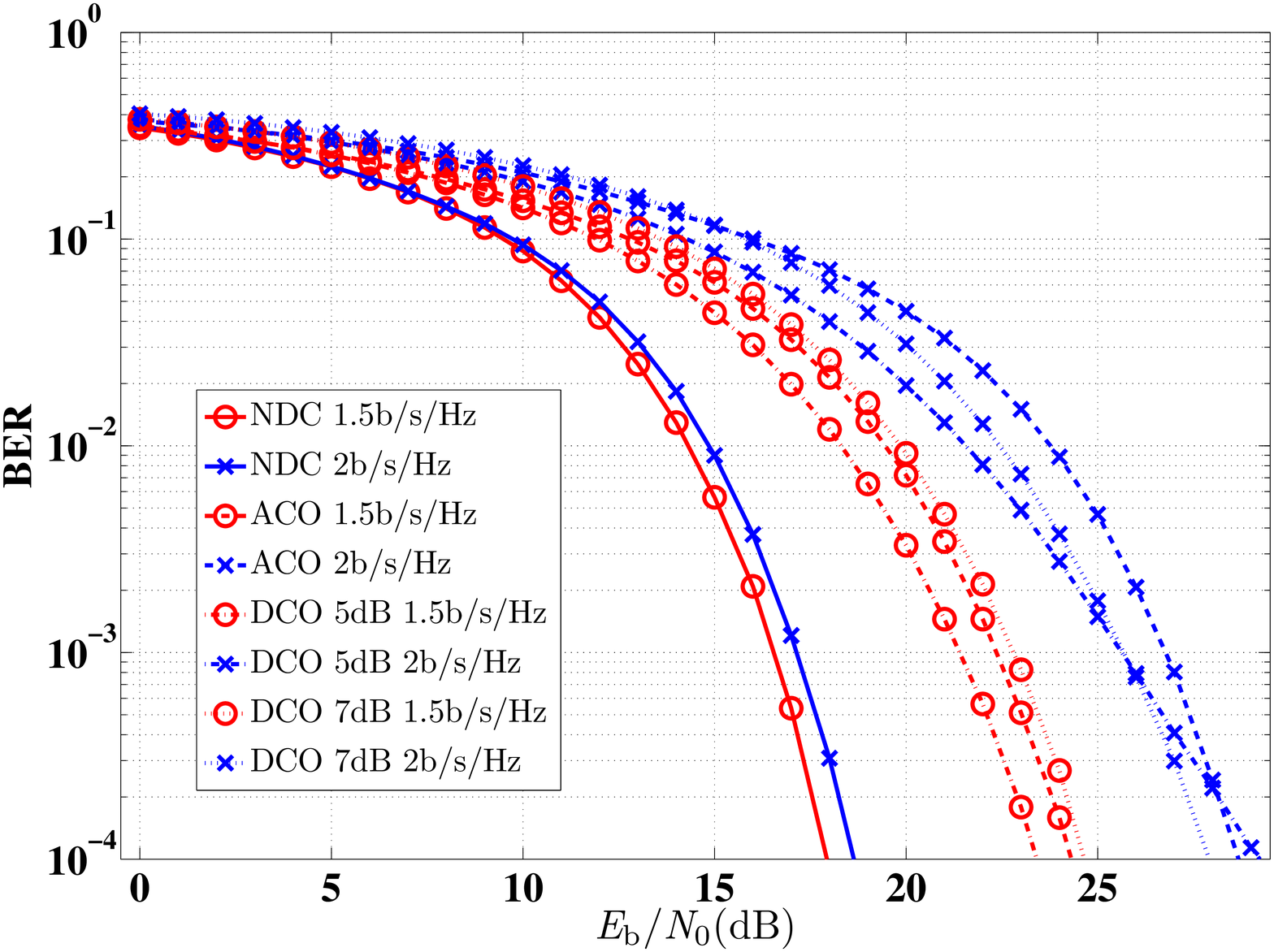}
\caption{NDC-OFDM, ACO-OFDM and DCO-OFDM Performance Comparison over ${\textbf H_{\rm Prac_4}}$}
\end{center}
\label{Hprac4}
\end{figure}

\section{Conclusions}

In this paper, theoretical and practical performances of a novel unipolar modulation method, called NDC-OFDM, is analysed. The new method combines O-OFDM with SM and has been applied to the OWC system. Due to the Bussgang theorem and the CLT, the closed-form analytical performance of NDC-OFDM in AWGN channels has been derived. As a result, an equation for the electrical SNR per bit, which presents a memoryless nonlinear distortion analysis processing, has been presented to calculate the theoretical BER performance of NDC-OFDM. In both ideal and practical channels, the derived BER result matches the Monte Carlo numerical result closely.

In comparisons of simulation performances, NDC-OFDM exhibits the capability for achieving higher energy efficiency than the conventional OFDM-based modulation schemes applied in the OSM system: DCO-OFDM and ACO-OFDM. Compared with DCO-OFDM, the new method solves the clipping distortion problem caused by the high level of the DC-bias. Additionally, it decreases the power consumption with less reduction in spectral efficiency as ACO-OFDM. Therefore, when VLC systems are developed towards equipped with multiple LEDs to fulfill minimum lighting conditions, NDC-OFDM will be used as an effective and efficient modulation scheme.

\section*{Acknowledgement}

\bibliographystyle{IEEEtran}
\bibliography{./cwc,./general}

\end{document}